\documentclass[12pt,dvips]{article}
\usepackage{graphicx}
\usepackage{amsmath}
\usepackage[psamsfonts]{amssymb}
\usepackage{amsthm}
\usepackage{euscript}

\usepackage{latexsym}

\newcommand{\feynslash}[1]{#1\hspace{-9pt}\slash\hspace{6pt}}
\renewcommand{\theequation}{\arabic{section}.\arabic{equation}}
\renewcommand{\(}{\begin{equation}}
\renewcommand{\)}{end{equation} \vspace{-.05in}\linebreak}

\newcounter{saveeqn}
\newcounter{savealpheqn}

\newcommand{\alpheqn}{\setcounter{saveeqn}{\value{equation}}%
 \stepcounter{saveeqn}\setcounter{equation}{0}%
 \renewcommand{\theequation}{\mbox{\arabic{section}.\arabic{saveeqn}\alph{equation}}}
 \renewcommand{\)}{\end{equation}}}
\def\part#1{\frac{\partial}{\partial{#1}}}%
\def\group#1{\refstepcounter{equation}\setcounter{saveeqn}{\value{equation}}%
 \label{#1}\setcounter{equation}{0}%
\renewcommand{\theequation}{\mbox{\arabic{section}.\arabic{saveeqn}\alph{equation}}}
 \renewcommand{\)}{\end{equation}}}
\newcommand{\reseteqn}{\setcounter{equation}{\value{saveeqn}}%
 \renewcommand{\theequation}{\arabic{section}.\arabic{equation}}%
 \renewcommand{\)}{\end{equation}}}

\newcommand{\aalpheqn}{\setcounter{saveeqn}{\value{equation}}%
 \stepcounter{saveeqn}\setcounter{equation}{0}%
 \renewcommand{\theequation}{\mbox{\Alph{subsection}.\arabic{saveeqn}\alph{equation}}}
  \renewcommand{\)}{\end{equation}}}
\newcommand{\areseteqn}{\setcounter{equation}{\value{saveeqn}}%
 \renewcommand{\theequation}{\Alph{subsection}.\arabic{equation}}%
 \renewcommand{\)}{\end{equation}}}

\renewcommand{\thefootnote}{\alph{footnote}}
\renewcommand{\(}{\begin{equation}}
\renewcommand{\)}{\end{equation}}
\newcommand{\ba}{\begin{eqnarray}}
\newcommand{\ea}{\end{eqnarray}}

\newcommand{\bp}{\mathop{\vtop{\ialign{##\crcr
  $\hfil\displaystyle{}\hfil$\crcr\noalign{\kern-13pt\nointerlineskip}
  \BIG{(}\hskip0pt\crcr\noalign{\kern3pt}}}}}
\newcommand{\cbp}{\mathop{\vtop{\ialign{##\crcr
  $\hfil\displaystyle{}\hfil$\crcr\noalign{\kern-13pt\nointerlineskip}
  \BIG{)}\hskip0pt\crcr\noalign{\kern3pt}}}}}
\newcommand{\pa}{\mathop{\vtop{\ialign{##\crcr
  $\hfil\displaystyle{\oplus}\hfil$\crcr\noalign{\kern+1pt\nointerlineskip}
  \hspace{.08in}$^{\alpha=0}$\hskip6pt\crcr\noalign{\kern3pt}}}}}
\renewcommand{\sp}{,\hspace{.3in}}
\newcommand{\p}{^\prime}
\newcommand{\pp}{^{\prime\prime}}

\newcommand{\R}{\ensuremath{\mathbb R}}

\newcommand{\Z}{\ensuremath{\mathbb Z}}

\newcommand{\del}{\ensuremath{\partial}}

\newcommand{\beq}{\begin{equation}}
\newcommand{\eeq}{\end{equation}}

\newcommand{\rp}{\mathbf{RP}}

\numberwithin{equation}{section}
\def\hsp#1{\hspace{#1in}}

\catcode`\@=11
\def\vereq#1#2{\lower3pt\vbox{\baselineskip1.5pt \lineskip1.5pt
\ialign{$\m@th#1\hfill##\hfil$\crcr#2\crcr\sim\crcr}}}
\catcode`\@=12

\makeatletter
\newcommand\figcaption{\def\@captype{figure}\caption}
\newcommand\tabcaption{\def\@captype{table}\caption}
\makeatother
\renewcommand{\(}{\begin{equation}}
\renewcommand{\)}{\end{equation}}
\newcommand{\etab}{{\overline{\eta}}}

\begin{document}

\begin{titlepage}
\begin{flushright}
hep-th/0210090
\end{flushright}

\vspace{2em}
\def\thefootnote{\fnsymbol{footnote}}

\begin{center}
{\Large SUSY vs $E_8$ Gauge Theory in 11 Dimensions}
\end{center}
\vspace{1em}

\begin{center}
Jarah Evslin\footnote{E-Mail: jarah@df.unipi.it}$^1$ and Hisham Sati\footnote{E-Mail: hsati@umich.edu}$^2$\ 
\end{center}

\begin{center}
\vspace{1em}
{\em $^1$INFN Sezione di Pisa\\
     Via Buonarroti, 2, Ed. C,\\
     56127 Pisa, Italy}\\
\hsp{.3}\\
{\em Department of Mathematics\\
     University of California\\
     Berkeley, CA 94720 USA}\\
\hsp{.3}\\
{\em $^2$Michigan Center for Theoretical Physics\\
     Randall Laboratory, University of Michigan\\
     Ann Arbor, MI 48109 USA}

\end{center}

\vspace{3em}
\begin{abstract}
\noindent

\end{abstract}
Diaconescu, Moore and Witten have shown that the topological part of the M-theory partition function 
is an invariant of an $E_8$ gauge bundle over the 11-dimensional bulk.  This presents a puzzle as an 11d 
gauge theory cannot exhibit linearly realized supersymmetry.  One possible resolution is that the gauge theory is nonsupersymmetric and flows to 11-d supergravity only in the infrared, with supersymmetry arising as a low energy accidental degeneracy.  Although no such gauge theory has been constructed, any such construction must satisfy a number of constraints in order to correctly reproduce the the known 10-dimensional physics on each boundary component.  We analyse these constraints and in particular use them to attempt an approximate construction of the 11d gravitino as a condensate of the gauge theory fields.

\vfill
October 10, 2002

\end{titlepage}
\setcounter{footnote}{0} 
\renewcommand{\thefootnote}{\arabic{footnote}}

\pagebreak
\renewcommand{\thepage}{\arabic{page}}
\pagebreak 

\section{The Motivation}

Six years ago Ho$\check{\textup{r}}$ava and Witten demonstrated \cite{HW2} 
that when M-theory is compactified on a manifold with boundary, 
the anomalies caused by chiral gauginos and gravitinos on each boundary component precisely 
cancel the anomalies that flow in from bulk. This cancellation 
occurs only if each boundary component supports precisely 248 10-dimensional vectormultiplets,
all transforming in the adjoint representation of $E_8$.  Furthermore as explained 
in Refs.~\cite{FluxQuant,DMW}, the topological
contribution to the M-theory partition function is in fact an invariant of the 
Dirac operator of a mysterious {\textit{11-dimensional}} $E_8$ gauge theory.

While the nature of this gauge theory is entirely unknown, the delicate anomaly-cancellation of Ho$\check{\textup{r}}$ava and Witten as well as the 10-dimensional $N=1$ supersymmetry on every boundary component place strong constraints on its construction.  We feel that the analysis of these constraints is a necessary first step in an attempt to understand the gauge theory.
  
In this paper we will 
propose the simplest possible particle content of such a proposal and then apply the above constraints.  In particular we will combine a constraint on the four-form, 10d SUSY covariance and 11d Lorentz covariance to construct the supergravity fields from the gauge fields.  We will then try to 
understand how this construction can be consistent with 
11d supersymmetry.  Such proposals have been considered previously in Refs.~\cite{Madrid, uday, allan, Morrison}. An apparently unrelated proposal which sacrifices the compactness of the $E_8$ but preserves supersymmetry has appeared in Refs.~\cite{NdW, Nic}.  While preserving the supersymmetry is an extraordinary advantage, it is not known whether the noncompactness of this $E_8$ would then lead to a noncompact gauge group for the heterotic string.  While providing a fascinating alternative to the class of gauge theories considered in this note, further speculation along these lines will be deferred to a sequel. 

For simplicity we will often restrict our attention to the case of flat, 
topologically trivial 11-dimensional space, although we generalize our 
results to curved space in Subsec.~\ref{CurveSec}.  We will also systematically neglect higher order Fermi
field contributions.  
  
It was shown in Ref.~\cite{HW} that $E_8$ gauge invariance combined with local supersymmetry invariance requires the relation 
\begin{equation}
\frac{G_4}{2\pi}=\frac{1}{16\pi^2}\textup{tr}(F\wedge F+\frac{1}{2}R\wedge R)
\label{g}
\end{equation}
between the 11d 4-form field strength $G_4$ and the 10D $N=1$ vectormultiplet's
fieldstrength $F$ on every 10-dimensional boundary component.  Following \cite{DMW} we consider an $E_8$ gauge bundle such that 
Eq.~(\ref{g}) holds everywhere in the 11-dimensional bulk.  The fact that such a bundle exists is a consequence of M-theory's shifted flux quantization condition \cite{FluxQuant}.  The uniqueness of this bundle results from the uniquely simple low dimensional topology of the $E_8$ group manifold.

In addition to the above 248 gauge bosons, we consider 248 adjoint Majorana fermions also propagating in the 11d 
bulk\footnote{And for now we also include an 11d graviton, although it is possible
that the graviton field is in fact a composite of the other gauge theory fields.}
.  Using Eq.~(\ref{g}) we can construct the 11d SUGRA 4-form $G_4$ from 
the vectors.  Ten dimensional $N=1$ SUSY covariance allows us to find the 
analogous construction of a chiral half of the 11-dimensional gravitino\footnote{Independent of conjectures about mysterious bulk gauge theories, we expect this relation of the gravitino to the gauge theory fields to hold on the boundary.} up to a mysterious problem related to the fact that we do not understand the role of the graviton in this story.
Eleven-dimensional Lorentz invariance allows us to construct the other half.
 Thus far, each gauge theory configuration is identified with a single 
SUGRA configuration, meaning that the construction 
cannot be covariant under 11d SUSY as the gauge fields are not part of any representation of 11d SUSY.

To remedy this we identify each gauge field configuration with not only the single SUGRA configuration given earlier, but with all of the SUGRA configurations which are related to that configuration by an 11d SUSY transformation.  Thus SUGRA 
field configurations related by SUSY transformations will be identified with the same gauge field configuration and thus the same 
physical state.  It is a critical check of the consistency of this construction that 
physically equivalent configurations on the gauge theory side are also equivalent on the SUGRA 
side, and in fact $E_8$ gauge transformations are realized as abelian gauge transformations of the M-theory 3-form.

In Sec.~\ref{setupsec} we review the standard arguments for an $E_8$ gauge theory in the bulk.  In Sec.~\ref{ConSec} we present our construction for the bulk gravitino in terms of gauge theory fields and show that this construction is consistent with 10 and 11-dimensional supersymmetries.  We conclude with some remarks on SUSY breaking, the graviton and also a relation to other $E_8$'s in the final section.

\section{$E_8$ Gauge Theory}\label{setupsec}

\subsection{Why an $E_8$ Bundle?}
The low energy effective description of M-theory is 11-dimensional supergravity \cite{CJS}.  The 
fields of this theory live in a single supermultiplet which contains the graviton, 
the gravitino $\psi$ and a three-form $C_3$ whose exterior derivative (times 6) is the 
four-form fieldstrength $G_4$.  If the dynamics of M-theory are to be formulated in terms 
of an $E_8$ gauge theory, it would be useful to have explicit relations between the fields 
of the 11d supermultiplet and the fields of the gauge theory: the 1-form connection $A$ with fieldstrength $F$ and an adjoint Majorana ``gaugino'' $\chi$.  

The conjectured relations arise from the synthesis of several observations.  First, in 
Ref.~\cite{HW2} it is shown that gauge and gravitational anomaly cancellation on any 
10-dimensional boundary of M-theory enforces the relation (\ref{g})
on the boundary, where $R$ is the curvature two-form (of the tangent bundle). In 
Ref.~\cite{FluxQuant} Witten used locality to argue that such relations, at the level of 
cohomology, can be extended to the bulk, although it does not follow from this argument 
that there is an $E_8$ gauge fieldstrength in the bulk.

One reason\footnote{Another very different reason has appeared in \cite{Morrison}.} that one may believe that there is in fact an $E_8$ gauge fieldstrength in the bulk 
is as follows.  The low energy effective action for M-theory on the 
11-fold $Y^{11}$ contains the topological terms
\begin{equation}
I=2\pi\int_{Y^{11}}C_3\wedge(G_4\wedge G_4-I_8) \label{i}
\end{equation}
where $I_8$ is a quartic in the curvature tensor.  Using a result from Ref.~\cite{HW} this can be related \cite{FluxQuant} to a sum of indices of an $E_8$ gauge theory on an auxilliary 12-dimensional manifold\footnote{More precisely, the ambiguity in $I$ is the integral of its exterior derivative over a closed 12-manifold.  This integral may be nonvanishing because $C_3$ is not necessarily globally defined.  The path integral measure is well defined if this integral, added to a contribution from the square root of the determinant of the Rarita-Schwinger operator, is an integer. It was shown in Ref.~\cite{HW} that the integral is in fact a sum of indices from an $E_8$ gauge theory and so is guaranteed to be an integer.}.  

In Ref.~\cite{DMW} a theorem of Atiyah, Patodi and Singer \cite{APS} was used to explicitly evaluate the contribution of this topological term and the Pfaffian determinant of the Rarita-Schwinger operator to the phase of the path integral measure:
\begin{equation}
\Phi=Pf(D_{RS})e^{i\int I}=|Pf(D_{RS})|{\textup{exp}}(\frac{2\pi i}{4}((h_{E_8}+\eta_{E_8})+ \frac{2\pi i}{8}(h_{RS}+\eta_{RS})).
\end{equation}
Here $\eta$ is the $\eta$-invariant of the corresponding operator (the $E_8$ gauge theory Dirac operator and then the Rarita-Schwinger operator) while $h$ is its number of zeromodes.  Thus a part of the path integral measure of 11-dimensional supergravity can be reexpressed in terms of a mysterious bulk $E_8$ gauge theory.  Furthermore it was shown that the partition function consists of a sum over $E_8$ gauge field configurations.

The fact that one factor in the M-theory partition function is the index of fermions 
charged under an $E_8$ gauge symmetry does not prove that there actually are fermions 
charged under an $E_8$, but the goal of the present paper is to understand how the 
existence of such fermions, and such a gauge symmetry, could be consistent with what
 we know of SUSY in 11 dimensions.

If there is such an $E_8$ gauge symmetry in the 11-dimensional bulk, a natural guess 
for its relation to the four-form fieldstrength is simply Eq.~(\ref{g}).
The rest of this paper will be an investigation of the consequences of this guess. The 
corresponding relation between the gravitino and gauginos will appear in Sec.~\ref{ConSec}.

\subsection{$E_8$ Bundles and Solitons}

If there is such a gauge theoretic description of low energy M-theory, it must be shown that $E_8$ gauge theory correctly reproduces the M-theory soliton spectrum \cite{allan}.  To compute the gauge theory's soliton spectrum we will need to review the topology of the group manifold $E_8$.

The low dimensional topology of $E_8$ is in one way the simplest among nonabelian Lie 
groups.  $E_8$\ has only one nontrivial homotopy group of dimension less than 15,
 which is $\pi_3(E_8)=\Z$.  This means that on a manifold of dimension less 
than 16, $E_8$ bundles are topologically characterized by a single characteristic class, 
the first Pontrjagin class 
\begin{equation}
p_1=\frac{Tr(F\wedge F)}{8\pi^2}. \label{p1}
\end{equation}
The only restriction on this class is that its integral over any 4-cycle be an even integer.
 All other semisimple Lie groups have additional nontrivial low dimensional homotopy groups and therefore their principal bundles cannot be completely characterized by a single characteristic class.
                                                                               
This agrees beautifully with what we know of M-theory, which at low energies is also 
described by a 4-form.  In fact substituting (\ref{p1}) into (\ref{g}) we learn that the 4-form flux of M-theory 
is a combination of this characteristic class and the first Pontrjagin class of the 
tangent bundle
\begin{equation}
\frac{G_4}{2\pi}=\frac{p_1(E_8)}{2}+\frac{p_1(TM)}{4}.
\end{equation}
Notice that the shifted flux quantization condition \cite{FluxQuant} of $G_4$ is automatic in 
this construction.  The first term on the right hand side is an integral cohomology class, 
while the second may be an integral cohomology class or may be half\footnote{By half of an 
integral cohomology class $\omega$ we mean consider the image of $\omega$ in the cohomology 
map induced by multiplication of the coefficient ring by 2 and then divide the answer by two.  
If there is $Z_{2k}$ torsion, then division by 2 is not well defined and so one needs a 
prescription for which quotient to take.  It was conjectured in \cite{FluxQuant} that the 
correct prescription is to force the answer to agree with the 4th Stieffel-Whitney class.} 
of an integral cohomology class.  Therefore the failure of the left hand side to be integral 
is precisely equal to the failure of the second term on the right hand side, that is, the $mod$\ 2 part of $p_1(TM)/2$.

As a result of the fact that an $E_8$ bundle is described by a single closed form, an $E_8$ bundle on a manifold of  
of dimension less than 16 has only one kind of topological defect, the 
M5-brane\footnote{If the soliton spectrum contains the M5-brane then it automatically contains the M2-brane.  For example, an M2-brane is created when two M5-branes cross via the Hanany-Witten mechanism \cite{HaW}, the M5-branes can usually be moved off to infinity.  Alternately, an M2-brane may be constructed as a limit of M5-branes that wrap a trivial 3-cycle supporting $C$ flux as integrated on a coordinate patch that contains the entire trivial cycle.   Such M5-branes are dielectric M2-branes \cite{Myers} and as the three-cycle shrinks to zero size become ordinary M2-branes.}.  This is the codimension 5 defect where the form fails to be closed.  String theory on backgrounds in which such a defect 
is linked by either a 4-sphere or $\rp^4$ have been studied extensively and in particular their soliton spectra are known.  We will now recover the 5-brane spectrum as the spectrum of $E_8$ defects. 

We will classify these defects by the restriction of their $E_8$ bundles to the 4-manifolds that link them. If this link is an $S^4$ then the bundle can be trivialized on the northern 
and southern hemispheres and the transition function on the 3-sphere equator must 
be an element $n\in\pi_3(E_8)=\Z$.  Likewise the bundle can be trivialized on the only 
hemisphere of an $\rp^4$ and the transition maps its equatorial $\rp^3$ to $E_8$.  These maps can be constructed by considering maps from $S^3$ to $E_8$, which are classified by $\pi_3(E_8)=\Z$, and then filtering them through $\rp^3$.  Only the maps corresponding to even integers can be filtered through $\rp^3$ and so the maps from $\rp^3$ to $E_8$ are classified by even elements $2n\in2\pi_3(E_8)=2\Z$.  Integrating first Pontrjagin classes over these two spaces one finds
\alpheqn
\begin{equation}
\int_{S^4}\frac{G_4}{2\pi}=
\int_{S^4}\frac{p_1(\textup{E}_8)}{2}+\int_{S^4}\frac{p_1(\textup{TM})}{4}=
\frac{2n}{2}+\frac{0}{4}=n
\end{equation}
\begin{equation}
\int_{\rp^4}\frac{G_4}{2\pi}=
\int_{\rp^4}\frac{p_1(\textup{E}_8)}{2}+\int_{\rp^4}\frac{p_1(\textup{TM})}{4}=
\frac{4n/2}{2}+\frac{2}{4}=n+1/2. 
\end{equation}
\reseteqn
By Gauss' Law these integrals are equal to the total M5-brane charge linked by the 4-cycle over which the integral is performed.  Thus the first configuration describes $n$ M5-branes, while the second describes an OM5 plane which carries $n+1/2$ units of M5-brane charge.  Recalling \cite{Kentaro} that OM5 planes always carry half integer charge we see that the spectrum of M5-brane charges is correctly reproduced by $E_8$ gauge theory.

\subsection{Analogy: The 'tHooft-Polyakov Monopole}
To gain some intuition for the construction (\ref{g}) of the four-form and for the role of the M5-brane defect we will, as suggested by Ref.~\cite{Harvey}, make a brief digression to consider a simpler system which shares many common features.  
Consider an SU(2) gauge theory in at least 3 dimensions with a scalar $\Phi^a$ that transforms in the adjoint of SU(2) and is subject to the potential
\begin{equation}
V(\Phi)=(1-\Phi^a\Phi_a)^2.
\end{equation}
The group SU(2), like $E_8$, is a simple Lie group and so 
$\pi_3(SU(2))=\Z$.  
Therefore the gauge bundle admits a codimension 5 defect constructed as the M5-brane is constructed above.  However, because a configuration of $\Phi$ is a map from spacetime to SU(2), the presence of an adjoint Higgs field in this model allows a defect of codimension 3.  If we impose a finite energy condition on 3-dimensional slices of spacetime then on each of these slices $\Phi$ is a map from $S^3$ times an irrelevant space to SU(2)$\cong S^3$.  Such a map is classified up to homotopy by an element $n\in \pi_3(\textup{SU(2)})=\Z$.  $n$ is the 'tHooft-Polyakov magnetic monopole charge of the 3-dimensional slice.

The fieldstrength $F$ of a U(1) gauge theory can be constructed from the 
fieldstrength $G^a$ of the original SU(2) and the scalar Higgs via
\begin{equation}
F=Tr(\Phi G)+\textup{$G$-indep}. \label{f}
\end{equation}
In the U(1) gauge theoretic description the 'tHooft-Polyakov monopole appears to be a Dirac monopole in the following sense:
\begin{equation}
\int_{S^2}F=1
\end{equation}
for any 2-sphere that links the monopole once.  However at microscopic distances from the core this abelian effective description breaks down and the physics (like asymptotic freedom) cannot be understood without knowledge of the full nonabelian model.  

This construction\footnote{Although in the $E_8$ case not all of the nonabelian DOFs can be encoded in the abelian fieldstrength.} could be repeated with an $E_8$ gauge theory that has a scalar Higgs transforming in the adjoint of the $E_8$.  In this case the configuration of the Higgs field in a 3-plane transverse to the monopole, after a 1-point compactification of this 3-plane, will again be an element $n\in\pi_3(E_8)=\Z$ where $n$ is the monopole charge.   

\subsection{Can an Abelian Four-form Describe a Nonabelian Theory?}

The existence of the adjoint scalar was crucial to the construction of the 
'tHooft-Polyakov monopole above.  Such a scalar does not exist in the 
$E_8$ gauge theoretic model of M-theory, and so there is neither a 'tHooft-Polyakov 
monopole nor an abelian two-form fieldstrength constructed as in (\ref{f}).  

In trying to construct an $E_8$ scalar from the $E_8$ fieldstrength analogously to (\ref{f}), we observe that although there is no adjoint scalar with which to contract its $E_8$ indices, there is the $E_8$ fieldstrength itself.  More precisely, we can construct an abelian four-form $\tilde{G}_4$ from the $E_8$ fieldstrength via
\begin{equation}   
\tilde{G}_4=\frac{1}{8\pi}\textup{tr}(F\wedge F)  \label{g2}
\end{equation}
so that $F$ also plays the role played by the Higgs field in the case of the monopole. As in the case of the 'tHooft-Polyakov monopole, this description breaks 
down in the core of a topological defect, where the nontriviality of $\pi_3(E_8)$ and thus the true nonabelian nature of the original high-energy theory become impossible to ignore.  

This breakdown of the abelian description near an M5-brane may have at least one consequence that has been observed in the literature.  In the M5-brane gravitational anomaly cancellation arguments of \cite{5Brane} there is a Dirac 
delta function in the modified Bianchi identity for $G_4$ which is very difficult to interpret when cubed in the 11-dimensional Chern-Simons action.  The attempt to interpret it in \cite{FHMM} has led to the introduction of an auxilliary bumpform, which may be explained by the behavior of fermion zeromodes near the core of the M5 \cite{HR}.  The interpretation in \cite{Lechner} again requires a ``bumpform'', the Chern kernel, whose exterior derivative is the distribution necessary to modify the SUGRA Bianchi identity while maintaining consistent couplings with M2-branes in the core of an M5.  However this argument relies on the existence of M2-branes as microscopic excitations of the theory.  In the $E_8$ gauge theory description the necessity of adding bumpforms is a symptom of using an abelian 4-form to describe a nonabelian theory in a region where the abelian description breaks down.  It may be interesting to compare the Chern kernel or the value of the bump calculated using the methods described in \cite{HR} with the result one would get starting in the nonabelian $E_8$ gauge theory and trying to ``approximate'' it by an abelian theory.

\section{The Construction} \label{ConSec}

\subsection{Constructing the Supergravity Fields}

Before relating the SUGRA and gauge theory fields, we will take a moment to review 10 and 11-dimensional SUGRA and to establish our conventions.  The 11-dimensional supermultiplet consists of an elfbein $e$, a gravitino $\Psi$ and a 3-form gauge potential $C$.
The transformations of these fields under 11-dimensional supersymmetry transformations are as follows \cite{CJS} 
\alpheqn
\begin{equation} \label{esusy}
\delta e_A{}^m={1\over 2}\bar\eta\Gamma^m\Psi_A
\end{equation}
\begin{equation} \label{csusy}
\delta C_{ABC}=-{\sqrt 2\over 8}\bar\eta  \Gamma_{[AB}\Psi_{C]}
\end{equation}
\begin{equation} \label{psisusy}
\delta\Psi_A=D_A\eta+{\sqrt 2\over 288}
\left(\Gamma_A{}^{BCDE} -8\delta_A^B\Gamma^{CDE}\right)\eta
G_{BCDE}
\end{equation}
\reseteqn
where $\eta$ is the 32-component Majorana spinor that parameterizes the 
variation.

The 10-dimensional vector supermultiplets \cite{HW} , which propagate on the boundary $M^{10}$,
consist of the $E_8$
gauge field $A$ (with field strength $F_{CD}=\partial_CA_D-\partial_DA_C+
[A_C,A_D]$) and spin $1/2$ Majorana-Weyl fermions (gluinos) $\chi$ in the adjoint representation,
obeying $\Gamma_{11}\chi=\chi$. Their supersymmetry transformation laws are
\alpheqn
\begin{equation} \label{axform}
\delta A_A^i = {1\over 2}\bar\eta\Gamma_A\chi^i
\end{equation}
\begin{equation} \label{chixform}
\delta \chi^i = -{1\over4}\Gamma^{AB}F_{AB}^i\eta
\end{equation}
\reseteqn
with spacetime indices $A,B=0,...,9$ in an orthonormal frame, and $E_8$ 
gauge group indices $i=1,...,248$, and
where $\eta$ is the 16-component Majorana-Weyl spinor that parameterizes the transformation.  

With these conventions established we may finally construct the gravitino from the gauge fields. 
It follows from the construction (\ref{g}) for the 4-form $G_4$ that for some $E_8$ gauge choice or equivalently
a gauge choice for $C_3$: 

\begin{equation}
C_{ABC}=\frac{1}{24 \pi} \big(
A_{[A}^i\partial_BA_{C]}^i+i\frac{2}{3}f^{ijk}A_{[A}^iA_B^jA_{C]}^k \big) . 
\label{ccon}
\end{equation}
Consider this relation restricted to a 10-dimensional boundary.  Although there is no bulk supersymmetry, the boundary theory is that of Ref.~\cite{HW} and so enjoys 10-dimensional $N=1$ 
supersymmetry.  Performing a rigid $N=1$ SUSY transformation on both sides we arrive at an expression for the gravitino

\begin{equation} 
\Gamma_{[AB}\Psi_{C]} =  - \frac{\sqrt{2}}{12 
\pi}\Gamma_{[A}F_{BC]}^i\chi^i.
\label{psicomposite}
\end{equation}  

where we have dropped pure gauge terms of the form $\partial_{B}(A_A 
\overline{\eta} \Gamma_C \chi)$. Contracting with $\Gamma^{ABC}$ gives 

\begin{equation}
{\feynslash{\Psi}}=\frac{\sqrt{2}}{12 \pi} \cdot \frac{1}{d-1}
\Gamma^{AB}F_{AB}^i \chi^i \label{slash}
\end{equation} 

On the other hand, contracting with $\Gamma^{AB}$, and using (\ref{slash}),
we get the fomula for $\Psi$ as
 
\begin{equation} \label{psicon}
\Psi_{C} = -\frac{\sqrt{2}}{12 \pi} \cdot \frac{1}{(d-1)} 
[a \Gamma^{AB}F^i_{AB}\Gamma_{C} + b \Gamma^{B}F^i_{BC}]\chi^i.
\end{equation}
where $a=\frac{1}{(d-2)}$ and $b=\frac{2(d-3)}{(d-2)}$.

This is a disturbing result, as a 10-dimensional supersymmetry transformation on 
the right hand side does \textit{not} yield (\ref{psisusy}).  First, it misses 
the derivative of $\eta$.  This is not a problem as we have only used rigid supersymmetry transformations and so this term vanishes.  However the variation of $\Psi$ produces 
an unwanted term, proportional to the kinetic term $F_{AB}F^{AB}$, that cannot 
so easily be dismissed. The origin of the term is as follows. The 
expression for $C_{ABC}$ is a wedge product 
of forms, but when we take the SUSY transformation on $A$ 
the one form changes to a zero form, $\chi$.  The SUSY variation of this $\chi$ yields an $F^2$ term that has two new dummy indices.  Due to the structure of the spinor spinor indices, these necessarily right-multiply the gamma matrices and so are not antisymmetrized with the indices of the other gamma matrix.  Thus $\delta \Psi$ contains a term with two $F$'s that are contracted, in stark contrast with the 11d SUSY transformation which yields a totally antisymmetrized four-form.

This may suggest that (\ref{ccon}) is not in fact 
SUSY covariant.  One might think this could be remedied by the addition of a closed form to the right hand side, however were such an addition required it would break the abelian gauge invariance of 11d SUGRA.  Instead one is therefore led to the conclusion that the original constraint, (\ref{g}), is not SUSY covariant, even on a 10d boundary.  Thus it appears as though another term would need to be added to the constraint to impose SUSY covariance.  The constraint on any such term is that it play the same role as the constraint in cancelling Eq. (2.9) of Ref.~\cite{HW2}.  Such an additional term, if we impose that it be $\Psi$-independent, appears not to exist.

One possible pessimistic conclusion is that the constraint is simply not at all SUSY covariant and so cannot be used to glean information about the gravitino via SUSY transformations.  However the problematic terms involve the graviton, whose role in this story and in particular whose relation to the gauge theory is as yet entirely mysterious.  Therefore one may interpret this apparent failure of covariance as a puzzle whose resolution places a very strong, if not lethal, constraint on the role that the graviton must play.  Another possibility is that the covariance has been destroyed by our truncations, and that were we to consider the curvature corrections and the higher order Fermi terms the covariance would be restored.  Below we will see that indeed the inclusion of curvature terms dramatically alters the form of this construction.

We are finally ready to extend our results to 11 dimensions.  We choose the gauge fields in the bulk so that these same relations hold.  However the 11-dimensional Lorentz group has no Majorana-Weyl representation, and so Lorentz invariance forces us to reinterpret the fermions in this construction as Majorana fermions. More generally there may be additional terms which vanish when the Weyl condition is imposed, to determine such terms one must impose SUGRA covariance.  To recover the above construction on the boundary, we impose the boundary conditions

\begin{equation}
A_{11}=0\sp \Gamma_{11} \chi = \chi
\end{equation}
where the $11$ direction is taken to be perpendicular to the boundary.

\subsection{SUSY Transformations}

So far we have a configuration of SUGRA fields $(C,\Psi)$ for every configuration of gauge fields $(A,\chi)$.  However our constructions (\ref{ccon}) and (\ref{psicon}) are not covariant under 11d SUSY transformations because the LHS transforms while the RHS is not in any 11d SUSY representation.  To attain covariance we will identify the entire gauge and SUSY orbit $(C\p,\Psi\p)\sim (C,\Psi)$ with each gauge orbit of $(A,\chi)$.  We will
now use the 11-dimensional SUSY transformations of the SUGRA fields to find constructions for all $(C\p,\Psi\p)$ related to the
unprimed fields by a single SUSY transformation with a small Majorana spinor parameter $\eta$.

For the three-form we obtain

\begin{eqnarray}
\frac{1}{24 \pi} \big(
A_{[A}^i\partial_BA_{C]}^i+i\frac{2}{3}f^{ijk}A_{[A}^iA_B^jA_{C]}^k \big)
&=&C_{ABC}=C\p_{ABC}-\delta_\eta 
C_{ABC}\nonumber\\
&=&C\p_{ABC}+\frac{\sqrt{2}}{8}{\etab}\Gamma_{[AB} \Psi_{C]}\nonumber\\
&=&C\p_{ABC}-\frac{1}{48 \pi} {\etab}\Gamma_{[A} F_{BC]}^i\chi^i
\end{eqnarray}
and similarly for the gravitino

\begin{eqnarray}
-\frac{\sqrt{2}}{12 \pi}\Gamma_{[A}F_{BC]}^i\chi^i
&=&\Gamma_{[AB}\Psi_{C]}=\Gamma_{[AB}\Psi\p_{C]}-\delta_\eta\Gamma_{[AB}\Psi_{C]}
\nonumber\\&=&\Gamma_{[AB}\Psi\p_{C]}-\Gamma_{[AB}D_{C]}\eta-\frac{\sqrt{2}}{288}\gamma_{ABC}^{DEFG}\eta 
G_{DEFG}\nonumber\\   
&=&\Gamma_{[AB}\Psi\p_{C]}-\Gamma_{[AB}D_{C]}\eta-\frac{\sqrt{2}}{288}\cdot\frac{1}{8 
\pi}
\gamma_{ABC}^{DEFG}\eta F_{DE}F^i_{FG}
\end{eqnarray}

where we have defined
\begin{equation}
\gamma_{ABC}^{DEFG}=\Gamma_{ABC}\Gamma^{DEFG}-8\delta_{[C}^D\Gamma_{AB]}\Gamma^{EFG}.
\end{equation}

Assembling these results the solution for  $(C\p,\Psi\p)$ is then
\alpheqn
\begin{equation}
C\p_{ABC}=\frac{1}{24 \pi} \big(
A_{[A}^i\partial_BA_{C]}^i+i\frac{2}{3}f^{ijk}A_{[A}^iA_B^jA_{C]}^k \big) 
+
\frac{1}{48 \pi}{\etab}\Gamma_{[A}F_{BC]}^i\chi^i
\end{equation}
\begin{equation}
\Gamma_{[AB}\Psi\p_{C]}= -\frac{\sqrt{2}}{12 \pi} 
\Gamma_{[A}F_{BC]}^i\chi^i
+\Gamma_{[AB}D_{C]}\eta
+\frac{\sqrt{2}}{288}\cdot\frac{1}{8 \pi}\gamma_{ABC}^{DEFG}\eta 
F^i_{DE}F^i_{FG}.
\end{equation}
\reseteqn

One can check that restricted to a 10-dimensional slice with $\eta$ Majorana-Weyl this transformation has the same effect on the 
right hand sides of the equations as a 10d SUSY transform on $A$ and $\chi$, and so reduces 
correctly to the 
Ho$\check{\textup{r}}$ava-Witten case.

By construction, the 11d supersymmetry algebra is still satisfied in this proposal.  For example, applying the above transformations twice on the 3-form $C$ one finds

\begin{eqnarray}
C\pp_{ABC}&=&C\p_{ABC}+\delta_{\eta\p}C\p_{ABC}=C\p_{ABC}+\delta_{\eta\p}C_{ABC}+\delta_{\eta\p}\delta_{\eta}C_{ABC}\nonumber\\
&=&\frac{1}{24 \pi} \big(A^i_{[A}\partial_B A_{C]}^i 
+i\frac{2}{3}f^{ijk}A^i_{[A}A^j_BA^k_{C]}\big)+\frac{1}{48 
\pi}(\overline{\eta}+\overline{\eta}\p)\Gamma_{[A}F_{BC]}^i\chi^i\nonumber\\
&& 
- 
\frac{\sqrt{2}}{8}\overline{\eta}\Gamma_{[AB}D_{C]}\eta\p
-\frac{1}{1152}\cdot \frac{1}{8 \pi}\overline{\eta}\gamma_{ABC}^{DEFG}\eta\p 
F^i_{DE}F^i_{FG}
\end{eqnarray}
and the second variation of $C$ is
\begin{equation}
\delta_{\eta\p}\delta_\eta C_{ABC}
= 
-\frac{\sqrt{2}}{8}
\overline{\eta}\Gamma_{[AB}D_{C]}\eta\p-\frac{1}{1152}\cdot \frac{1}{8 \pi}
\overline{\eta}\gamma_{ABC}^{DEFG}\eta\p F^i_{DE}F^i_{FG}.
\end{equation}
The commutator of two such variations is
\footnote{A useful property here is the Majorana flip in 11 
dimensions \cite{VN}

\begin{equation}
({\bar\lambda} \Gamma^{I_1} \Gamma^{I_2} \dots \Gamma^{I_n} \eta)=(-1)^n
({\bar\eta} \Gamma^{I_1} \Gamma^{I_2} \dots \Gamma^{I_n} \lambda) .\nonumber
\end{equation} }

\begin{equation}
[ \delta_{\eta\p},\delta_{\eta} ] C_{ABC}= -\frac{1}{3}
\overline{\eta}\Gamma^D \eta\p G_{ABCD}
\end{equation}
reproducing the Poincar\'e supersymmetry algebra in flat 
space \cite{CJS}, up to pure gauge terms of the form 
$\partial_{[A}\Lambda_{BC]}$, with 
$\Lambda_{BC}=C_{BCD}\overline{\eta} \Gamma^D \eta\p$. 

\subsection{Curvature Corrections} \label{CurveSec}

We now describe the full construction, including the term $Tr(R \wedge R)/2$ in $G_4$ although not including higher order fermion corrections.  We begin with the variation of the Riemann tensor with respect 
to the metric using the Palatini formula  \footnote{We denote the 
Christoffel symbols by ${\tilde \Gamma}$ to avoid confusion with the Gamma 
matrices.}
 \begin{equation}
\delta R^A{}_{BCD}=\nabla_{C}(\delta {\tilde \Gamma^A{}_{BD})}-
\nabla_{D}(\delta {\tilde \Gamma^A{}_{BC})}. 
\end{equation}
The variation of the Christoffel symbols with respect to the metric is
\begin{equation}
{\delta \tilde \Gamma^A{}_{BC}}=\frac{1}{2} g^{AE} [\del_C (\delta g_{BE})
+ \del_B (\delta g_{CE})- \del_E (\delta g_{BC}) ].
\end{equation}
Using the supersymmetry variation of the metric, one obtains 

\begin{equation}
\delta R^A{}_{BCD}=\frac{1}{2} g^{AE}\big[ 
\nabla_C [\del_D ({\bar \eta} \Gamma_{(B} \Psi_{E)})
          +\del_B ({\bar \eta} \Gamma_{(D} \Psi_{E)}) 
          -\del_E ({\bar \eta} \Gamma_{(B} \Psi_{D)})]
- [C \leftrightarrow D] \big]
\end{equation}
Finally we contract with another $R$ to find the desired
\begin{eqnarray} \label{drr}
\delta Tr(R \wedge R)= 
\big[ \nabla_C [\del_D ({\bar \eta} \Gamma_{(B} \Psi_{A)})
          +\del_B ({\bar \eta} \Gamma_{(D} \Psi_{A)})
          -\del_A ({\bar \eta} \Gamma_{(B} \Psi_{D)})]
- [C \leftrightarrow D] \big] R^{ABCD}
\end{eqnarray}
where it is understood that the indices of $\Gamma$ are symmetrized with
those of $\Psi$. The full construction of $\Gamma\Gamma\Psi$ is then 
simply the old construction plus an inverse exterior derivative of the right 
hand side of (\ref{drr}). 

More explicitly, the M-theory 3-form is

\begin{equation}
C_3= C.S.^{(gauge)} + C.S.^{(grav.)}
\end{equation}
where
\begin{equation}
C.S.^{(grav.)}=\frac{1}{48 \pi}\epsilon^{ABC}\big[
{\tilde \Gamma^D_{AE}} \del_B {\tilde \Gamma^E_{CD}} 
+ \frac{2}{3} {\tilde \Gamma^F_{AD}} {\tilde \Gamma^D_{BG}} 
{\tilde \Gamma^G_{CF}} 
\big] .
\end{equation}

The curvature-corrected construction of the gravitino is then

\begin{equation}
-{\sqrt 2\over 8}\bar\eta  \Gamma_{[AB}\Psi_{C]}=\frac{1}{48\pi}\Gamma_A F^i_{BC}\chi^i+\delta C.S.^{(grav.)}
\end{equation}

where the variation 
of the $C.S.^{(grav.)}$ is

\begin{eqnarray}
\delta C.S.^{(grav.)} = \frac{1}{48 \pi} \cdot
\frac{1}{4} \epsilon^{ABC} \big[ g^{DH}
  [ \del_E ({\bar \eta} \Gamma_{(A} \Psi_{H)}) 
    + \del_A ({\bar \eta} \Gamma_{(E} \Psi_{H)})
    - \del_H ({\bar \eta} \Gamma_{(A} \Psi_{E)})] \del_B 
{\tilde \Gamma^E_{CD}}
\nonumber\\
  + {\tilde \Gamma^D_{AE}} \del_B [g^{EH} 
  [ \del_C ({\bar \eta} \Gamma_{(D} \Psi_{H)})
  + \del_D ({\bar \eta} \Gamma_{(C} \Psi_{H)})  
  - \del_H ({\bar \eta} \Gamma_{(C} \Psi_{D)})]
\nonumber\\
  +{\frac{2}{3}} g^{FH}[ \del_A ({\bar \eta} \Gamma_{(D} \Psi_{H)})
  + \del_D ({\bar \eta} \Gamma_{(A} \Psi_{H)})  
  - \del_H ({\bar \eta} \Gamma_{(A} \Psi_{D)})]
   {\tilde \Gamma^D_{BG}} {\tilde \Gamma^G_{CF}}  
\nonumber\\  
  + {\frac{2}{3}} {\tilde \Gamma^F_{AD}} g^{DH}
[  \del_B ({\bar \eta} \Gamma_{(G} \Psi_{H)})
 + \del_G ({\bar \eta} \Gamma_{(B} \Psi_{H)})
 - \del_H ({\bar \eta} \Gamma_{(B} \Psi_{G)})] {\tilde \Gamma^G_{CF}}
\nonumber\\
+{\frac{2}{3}} {\tilde \Gamma^F_{AD}} {\tilde \Gamma^D_{BG}} g^{GH}
[ \del_C ({\bar \eta} \Gamma_{(F} \Psi_{H)})
 + \del_F ({\bar \eta} \Gamma_{(C} \Psi_{H)})
 - \del_H ({\bar \eta} \Gamma_{(C} \Psi_{F)})]
\big] .
\end{eqnarray}
This is a first order linear differential equation for the construction of the gravitino $\Psi$.  Notice that even on a 10-dimensional boundary, when curvature terms are not omitted, the 10-dimensional SUSY variation of (\ref{g}) yields only a differential equation for the restriction of $\Psi$ to the boundary because derivatives of $\Psi$ appear in the SUSY variation of the curvature.
   
The above 
analysis could have equivalently been done using vielbeins and spin connections.
Given an expression of the metric in terms of the gauge fields, an explicit 
one for the Christoffel symbols and the Riemann tensor 
could be constructed. This will be left to future work.





 
\section{Conclusion}

The M-Theory partition
 function was originally proved to be well defined \cite{FluxQuant} using a mysterious 
$E_8$ bundle which restricts from 12 dimensions to the 11-dimensional bulk.  We have 
tried to understand how the existence of such a bundle can
 be compatible with 11-dimensional supersymmetry.  We have considered a class of gauge theories that is not in fact supersymmetric, but rather SUSY appears as a low energy accidental symmetry.  At slightly higher energy scales one would therefore observe SUSY to be slightly broken.  It may be interesting to compare this pattern of supersymmetry breaking with phenomenology.

A natural
goal is to explain how, via integrating out loops, the
11d SUGRA
fields become the correct low energy DOFs of some $E_8$ gauge theory and in particular to reproduce
the 11d
SUSY action as a low energy effective action.  It was implicit
throughout this paper that the 11d SUGRA action was produced correctly and 
in particular that the SUGRA fields enjoy an 11d SUSY invariance.  Of
course,
figuring out what this gauge theory is will be the ultimate goal, and its
hoped that the fact
that it reduces in the IR to 11d SUGRA will be a crucial constraint,
hopefully not so powerful of a constraint that the gauge theory cannot exist.  Such a gauge theory is likely to require a UV completion and so will be at best only a low energy approximation to M-theory.

An analogous construction may exist for the elfbein or the graviton, that is the graviton may itself be a condensate of fields in the gauge theory.  This possibility is currently under investigation.  Another tantalizing venue of future investigation is to investigate the link between this $E_8$ and that of the 11d $E_{8(8)}$ SUGRA of Nicolai and de Wit \cite{NdW,Nic}.

\noindent 
{\bf Acknowledgements}

\noindent  
We would like to thank Allan Adams, Kurt Lechner, James Liu, Pieralberto Marchetti, 
Darius Sadri, Eric Sharpe, Uday Varadarajan and Jay Wacker for many useful 
questions and comments.  In particular we would like to thank Jim and Uday for finding 
an apocolyptic errors.  This work was made possible by the good people of the 
INFN and by DOE grants DE-FG02-95ER40899 and DE-FG02-95ER40899.

\noindent


\end{document}